# Interpreting the Caste-based Earning Gaps in the Indian Labour Market: Theil and Oaxaca Decomposition Analysis

Pallavi Gupta

(BSE Institute Ltd., Mumbai, INDIA)

Satyanarayan Kothe

(Mumbai School of Economics and Public Policy, University of Mumbai, Mumbai, INDIA)

## Abstract

The UN states that inequalities are determined along with income by other factors - gender, age, origin, ethnicity, disability, sexual orientation, class, and religion. India, since the ancient period, has socio-political stratification that induced socio-economic inequality and continued till now. There have been attempts to reduce socio-economic inequality through policy interventions since the first plan, still there are evidences of social and economic discrimination. This paper examines earning gaps between the forward castes and the traditionally disadvantaged caste workers in the Indian labour market using two distinct estimation methods. First, we interpret the inequality indicator of the Theil index and decompose Theil to show within and between-group inequalities. Second, a Threefold Oaxaca Decomposition is employed to break the earnings differentials into components of endowment, coefficient and interaction. Earnings gaps are examined separately in urban and rural divisions. Within-group, inequalities are found larger than between groups across variables; with a higher overall inequality for forward castes. A high endowment is observed which implies pre-market discrimination in human capital investment such as nutrition and education. Policymakers should first invest in basic quality education and simultaneously expand post-graduate diploma opportunities, subsequently increasing the participation in the labour force for the traditionally disadvantaged in disciplines and occupations where the forward castes have long dominated.

# 1. Introduction

India, since the ancient period, experienced socio-political stratification that induced socio-economic inequality and continued till now. There have been attempts to reduce socio-economic inequality through policy interventions since the first plan. Despite the phenomenal economic growth in India in recent years, the UN reports that income inequality has increased in most developed countries including China and India since 1990 (UN, 2020). It further states that inequalities are determined along with income by other factors - gender, age, origin, ethnicity, disability, sexual orientation, class, and religion; these factors determine the inequalities of opportunities within and between countries (UN, 2020). The division of society into unequal and hierarchical social categories by limiting access to resources or services related to education, health, jobs, assets such as land and even exclusion from public participation and decision making has been the primary source of social stratification and has caused inequalities in opportunities and outcomes for those considered as 'lesser' or 'impure' to the traditionally advantaged groups (Das & Dutta, 2007). Such discrimination has implications in the labour market given that such divisions are rooted in the occupational division of labour with the lower or marginalised castes subject to low end, menial and manual work considered 'most degrading and ritually unclean' such as sweeping and leather work (Gang et al., 2012). In India, this peculiar concept of societal hierarchy has manifested itself into sharp contrast in incomes between the scheduled castes (SCs), scheduled tribes (STs) and other backward castes (OBCs) on one hand and the traditionally advantaged so-called forward castes (FCs) on the other. The first three: SCs, STs and OBCs are classified as underprivileged/ marginalised groups while the FCs are classified as a privileged group or high caste. Historically, the SCs are the untouchable group and have socially been placed outside of society for centuries. The second group, STs, are a group whose distinction is made based on language and cultural activities from the rest of Indian society. These two groups were often not allowed to participate in most of the economic decision-making process in India. The OBCs were classified as those who were not part of the former two groups and neither were part of the upper caste. This group is also deprived, both socially and economically. These four groups represent approximately 20, 8, 42 and 30 per cent of the population of India respectively (Reza et al., 2018).

Even among those traditionally disadvantaged groups, "within all scheduled castes some sub-castes would be more dominant while sub-castes like *chamars* are unambiguously at the bottom of the societal hierarchy" (Das & Dutta 2007), therefore the scheduled castes have a higher incidence of poverty attributed to unequal treatment, higher incidence of unemployment, and earning significantly less than the non-scheduled castes (Deshpande & Sharma, 2016; Kijima, 2006). Literature has pointed towards discrimination of scheduled caste workers during hiring for example due to restrictions on taking up jobs traditionally done by higher caste workers, therefore, resulting in non-voluntary unemployment (Thorat, 2018). There are difficulties attached to attaining regular employment and labour market entry and even for those employed, the work conditions are associated with low payments and worse working conditions without any social security. The last of the National Sample Survey (2011-12) documents that as compared to the forward caste workers, the SCs are employed for a lesser number of days; with an average annual loss of employment of about 28 days. It shows the proportion of wage labourers is highest for the SCs at 63 per cent and even the unemployment rate among the SCs is 1.7 per cent more than the others. Studies have documented the extent of caste-based discrimination. A survey conducted during 1993 in four different states of India, showed that about 41 per cent of farm wage labourers were denied work such as harvesting and cultivation due to prejudice attached to lower castes (IIDS, 1993). In urban areas, a different study documented a similar prejudice faced in hiring; with scheduled castes having 67 per cent less chance of being called for an interview despite being equally qualified as the forward caste applicant (Attewell & Thorat, 2010).

It is seen that such contrast has not only manifested in occupational divisions as a reflection of a highly unequal labour market but also largely due to pre-existing 'visible' discriminations in access to education and educational attainment (Borooah & Iyer, 2007) and unequal wages working in same jobs (Banerjee & Knight, 1985; Das & Dutta, 2007) and 'implicit' ones as attitudes and behaviour of the employers (Gang et al., 2012) that elevates income disparities between the socially disadvantaged and forward group workers. Discrimination, therefore, fabricates a long run disadvantaged condition and limits the opportunities for one group to another (Bourguignon et al., 2007). There is little doubt that regardless of the form of discrimination, whether based on caste, race, gender or skin colour it does lead to significant gaps in earning opportunities and wages (Esteve-Volart, 2009; Klasen, 2018).

In the context of existing social discrimination, examining earnings gaps between castes can give insight into the core of the inequality problem. The objective of this paper, therefore, is to examine the earnings differentials between the forward castes and the traditionally disadvantaged groups such as the SCs, STs, and OBCs separately for the urban and rural divisions. We focus on using two different empirical approaches to achieve our objective. First is the Theil index. Theil allows us to examine aggregate disparities in earnings between castes, and we do this separately for region (rural/urban), sector (public/private) and gender (female/male) classifications. Subsequently, we decompose Theil into within and between groups allowing us to see which of the two is mostly causing such gaps. A more accurate method is through examining earnings gaps based on observable characteristics between groups is the Blinder-Oaxaca decomposition (Blinder & Oaxaca 1973). The methodology is widely used to understand labour market outcomes by groups such as gender, caste, region and so on where mean differences in log wages are decomposed following a regression model (Jann & Zurich, 2008, p. 2). Here, the way to interpret earning differentials is to decompose the observed gaps into i) endowments or 'observable characteristics'; ii) coefficient or 'unexplained component; and iii) interaction or part of the indirect effect that can be further decomposed into a 'pure indirect effect and meditated interactive effect' (Cotton 1988; Newmark 1988). This method is commonly applied in the Indian labour market and to study discrimination based on caste and/or gender segregations (Poddar & Mukhopadhyay, 2019; Das, 2018; Lama & Majumdar, 2018; Agrawal, 2013 among others).

The paper is organised as follows. Section 2, reviews the literature. In Section 3, the choice of data sources and methodology is mentioned. We also present the data description, number of variables and the use of dummy variables in Section 4. Subsequently, we explain the Theil Index and its decomposition along with the Oaxaca Decomposition method in Section 5. The empirical results are mentioned under Section 6. Section 7 concludes with discussion and policy implications.

## 2. Literature Review

The role of caste-based discrimination in the Indian labour market is well documented by Madheswaran & Singhari (2016) describing significantly lower wages for SCs in both public and private sector and wage differentials attributable to occupational discrimination which means discrimination in access to employment; rather than discrimination within an occupation. In occupations witnessing the largest increases in wages (sectors such as IT and management), the share of marginalised workers is underrepresented, resulting in greater wealth concentration among forward castes in the recent decades (Bharti, 2018).

While there has been extensive research on caste and wealth inequality, most papers have focused on differences in inheritance, access to education and discrimination (Borooah, 2005; Tagade et al., 2018). Zacharias and Vakulabharanam, 2011 highlight that economic differences among castes precisely conform to the caste hierarchy present in society. This suggests the existence of caste-based 'group inequality' in India, a concept developed in Jayadev and Reddy (2011) to measure within inequality between groups in a population. Findings from IHDS data support this theory, showing that group indicators such as average skill-level, wealth and consumption are all ordered hierarchically along caste lines (Bharti, 2018). Lower educational attainment of backward castes means that they are differentially impacted by a loss of middle-skilled jobs, and the consequent inability to find work may make it harder for them to accumulate wealth and skills, thus aggravating inequality between castes. There seems to be a correlation between wage and education level of individuals from different socio-religious groups (Agarwal, 2013; Kingdon, 1998) where the focus is on the effect of differential returns to education to explain labour market outcomes. The differences in the quality and amount of education are thus understood as the main factors determining one's ability to secure regular salaried employment among the traditionally socio-disadvantaged classes.

NSS-EUS has been utilised in literature to estimate the degree of caste-based inequality in wages while considering demographic differences between groups. It is seen that within 'different age cohorts' of forward caste and traditionally disadvantaged groups, income gaps are increasing (Arabsheibani et al., 2018) and such gaps among workers with higher education are attributed to the prevalence of discrimination that results in lower wages (Madheswaran and Attewell, 2007). Sidkar (2019) attempts to determine earning differentials between the "formal and informal" sector for socio-religious group individuals, showing a significant relationship in case of socio-classified groups between wages and education levels, while considering all educational levels, and yet persons belonging to general category with higher educational level are able to get better jobs, none of the other three groups, i.e., the SCs/STs and OBCs seem to show any substantial impact of higher education on wages.

Although caste is principally an Indian phenomenon, its effect on wages is studied in other countries like Nepal and Bangladesh using Blinder Oaxaca decomposition techniques to find large wage differentials. Mainali et al. (2016) uses Blinder-Oaxaca decomposition techniques to find that wage-differentials due to caste are large, though this caste-based wage-differential is caused by the difference in investments in "human capital" and also low opportunities for "high-paying" jobs. Their decomposition method is expanded to consider firm sizes, and they also find that underrepresentation of lower castes in larger firms contributes significantly to the overall wage differential. Karki & Bohara (2014) conduct a Blinder-Oaxaca as well as a non-parametric decomposition analysis on monthly earnings data from Nepal and find that "differences in endowment" causes significant gaps in wages between Dalits and non-Dalits. Decomposition analysis conducted on data from Bangladesh also suggests the existence of a strong gender-based sticky floor effect i.e., wage-differentials due to discrimination are highest among low quintiles of the income distribution, and a "weaker glass-ceiling effect" (Faruk, 2019).

Within the Indian context, Blinder-Oaxaca decomposition has been employed to understand caste-based differences in various contexts. Sangwan (2020) conducted a decomposition analysis on India Human Development Survey data from 2005 to 2011-12 to find whether credit access varied based on caste. Substantial evidence for caste-based differences in credit access is found after correcting for selection bias. Bhuyan et al. (2018) used Oaxaca-quantile decomposition techniques to analyse differences in food security of backward and forward castes in both rural and urban India. Unsurprisingly, they found that the incidence of food insecurity was higher among lower castes, though more of this differential was explained through differences in overall identity than caste. Kumar and Pandey (2021), have explained the factors contributing to large discrimination caused by lack of formal employment in India using a three-fold Blinder Oaxaca-decomposition method.

Some recent studies reinstate the continuing influence of caste on wealth inequality. Thorat and Madheswaran (2018) find asset ownership differences to be the most enduring source of caste-based inequality in consumption spending (followed by differences in educational qualifications). Importantly, they find that the magnitude of the wage differential as a consequence of caste varies across the wage distribution: it is higher in the upper quintiles and lowers among the bottom quintiles. This is in contradiction to Mainali et al. (2016) who finds that in Nepal, the greatest discrimination occurs at the lowest quintiles of the wage distribution. In agreement with Mainali et al. (2016), using Oaxaca decomposition techniques, Khanna (2012) finds a higher gender wage differential among lower quintiles of the wage distribution when compared to the uppermost quintiles. The impact of reservation policies on income inequality between castes is studied in Brennan et al. (2006), however, there are fewer attempts to understand the implications of discrimination causing a persistent wage inequality among different castes. This paper attempts to fill the gap, by incorporating rural and urban populations and also studying discrimination within different occupational divisions.

## 3. Data Sources and Methodology

For this paper, a micro-individual data file for the 68th round (2011-12) is accessed to calculate both Theil index and its decomposition, and subsequently Three-Fold Oaxaca decomposition; the detailed explanation for both is done subsequently under this section.

For our analysis, we include four social classifications: scheduled tribes (STs), scheduled castes (SCs), other backward categories (OBCs) and 'others' or general category workers. For notification purposes, we use FC for all forward caste (referred to as "Others") workers, and NFC for all those belonging to backward caste (including SCs, STs, OBCs). We include both rural and urban workers in the regular salaried category of the Indian labour market.

Data for wages in NSS-EUS is available for employed individuals as regular salaried and casual workers. Wages are given in rupees as 'received or receivable' on a weekly work done basis. For analysis purposes, we focus on wages paid in cash and kind and convert wage and salary earnings that are given as current weekly status (CWS) to a 'daily rate'. The daily rate[1] is thus derived as a ratio of the given weekly wage and the number of either half-day or full-day work for the given week. We exclude exclusive gender and religious segregation from this analysis. For occupations, Broad occupational divisions as per NCO 2004 is considered[2].

In the context of widening gaps in incomes and opportunities, we interpret earning differentials (and its components) between the forward caste workers (FC) and the non-forward caste workers (NFC). We use two different approaches in this paper to analyse discrimination in wages and occupations. First, a common measure used for evaluating income inequality is the Theil index that was originally proposed by Theil (1967) and is widely applied to social and economic sciences mostly because of its decomposability (Liao, 2016; Akita 2003; Allison, 1978). For this paper, we calculate the Theil index using the STATA software's inbuilt command "ineqdeco" following the

---

[1] Accordingly, the NSS survey considers 'full day' if an individual works on any activity in one day for four hours or more and it considers 'half day', if work is between one to four hours in one day.
[2] See Notes for Occupational Divisions; NCO 2004.

Stephen P. Jenkins model of Theil Index (Jenkins, 1999)[3]. We utilise the Theil index to calculate aggregate income differentials between the *FC* and the *NFC* based on gender, sector and region. Subsequently, we decompose the Theil index into 'within' and 'between' components. Decomposing Theil Index is helpful since it allows us to show the extent of wage discrimination prevalent between non-forward caste and forward caste workers belonging to the same group (Liao, 2016). Previous literature has documented the role of 'within groups' to be a significant contributor to overall inequality in India. Theil also incorporates decomposition for multilevel data such as that provided under NSSO unit record (micro-level) data. A refined methodology as explained in Liao (2019) takes into account 'glass-ceiling' and 'glass-floor effects' that further decomposes within-group inequality in two subcomponents. We leave this formulation outside the scope of this paper.

Our second approach is to employ the Oaxaca Decomposition technique. We use the STATA inbuilt command (Oaxaca) developed by Ben Jann (Jann, 2008; Jann & Zurich, 2008). This method allows us to essentially segregate the differences in mean wages into "endowment" and "coefficient components". Differences in productivity variables represent differences in wages due to skill, whereas differences in coefficients represent potential discrimination. Mincerian Earnings Function (Mincer, 1974) is calculated to study the impact of education and experience on wages. OLS regression is run separately for the four divisions of castes: SCs, STs, OBCs and others. Estimation of returns to schooling is a crucial piece in completing the study of unequal distribution of wages across the population since it helps in shedding light on why certain groups may remain disadvantaged in terms of earnings, and why education does improve earnings. Estimating Returns on education brings out the level of discrimination and inequality faced by the disadvantaged groups at each level of education. It also helps us analyse which group of the four divisions face more barriers (and whether they do at all) in climbing the education level.

**4. Number of Observations, use of dependent, independent and dummy variables**

---

[3] Ineqdeco estimates different inequality indices that are commonly used by researchers. It also allows decompositions by subgroups that are helpful in estimating inequality profiles at a point in time, and shift-share analysis. Ineqdeco estimates inequality indices that belong to single parameter Generalised Entropy class GE (a) for a = -1, 0, 1, 2; the Atkinson class A(e) for e = 0.5, 1, 2; the Gini Coefficient, and the percentile ratios p90/ p10/ and p75/ p25. The more positive the inequality indices, GE (a) is more sensitive to income difference at the top of the distribution, the more negative inequality indices reflect greater sensitivity to difference at the bottom of the distribution. GE (0) is the mean logarithmic deviation, GE (1) is the Theil index, and GE (2) is the half the square coefficient of the variation (also see Stephen P. Jenkins (2008) for detailed explanation of ineqdeco).

After filtering data from the unit level data files of NSSO, the total number of observations are 70,067 individuals, of which 20,125 are those belonging to $FC$ and 49,942 are those belonging to the $NFC$. We take the value of the dependent variable of probit (selection) as 1 if an individual wage is > 0, and 0 otherwise. Therefore, we include workers with non-zero income in the age bracket of 15 to 60 years. The workers belonging to the regular salaried and casual labour market are considered.

We take the dependent variable (outcome of interest) to be the natural log of daily wage. Variables of age, levels of education, region, occupation and industry are taken as predictors. The data does not provide years of work experience; therefore, we use age as an approximation to experience. We use different dummy variables for controlling the household characteristics such as gender (male/ female), type of employment (regular/ casual) and sector (rural/ urban) to get a better estimate for establishing a relationship between education and wages for each caste category separately (Please see Appendix 1 for details).

A previous round of NSS (64th round), documents participation and expenditure in education along with the years of formal schooling among the population covered in the survey between age 5 to 29 years. The level of 'general education' provides the maximum level of education completed, which is similar to the NSS-EUS. *Codes*[4] assigned for all levels of education are as follows: "primary (06)", "middle (07)", "secondary (08)", "higher secondary (10)", "diploma/ certificate course (11)", "graduate (12)" and "postgraduate and above (13)".

### 5a. Explanation of Theil's Index and its Decomposition Method

First, total inequality as measured by Theil is given as:

$$T = \frac{1}{N} \sum_{j=1}^{N} \frac{x_j}{\bar{x}} \ln \frac{x_j}{\bar{x}} \qquad 1$$

Here $j = 1, 2, 3, \ldots, n$ with $x_j$ = individual's ($j$) income, $\bar{x}$ = mean income and $N$ = size of the population.

Equation (1) above can be additively decomposed into two parts:

---

[4] Please note that codes assigned are not the same as "average years of education".

$$T_b = \sum_{k=1}^{K} y_k \ln \frac{\bar{x}_k}{\bar{x}} \qquad 2$$

First being the "between group" inequality here, $y_k$ = subgroup $K$'s income shares as a proportion of the total income of full sample, $\bar{x}_k$ = group $k$'s mean income.

$$T_w = \sum_{k=1}^{K} y_k \sum_{j=1}^{n_k} y_{jk} \ln \frac{x_{jk}}{\bar{x}_k} \qquad 3$$

The second being the "within-group" inequality. Here, $jk$ = income share of an individual ($j$) within subgroup $k$, and $x_{jk}$ = individual $j$'s income within subgroup $k$.

### 5b. Explanation of Three-fold Blinder Oaxaca Decomposition Method

First taking the gross wage differential (denoted as $W$ between the $FC$ and the $NFC$ groups is the difference in the predicted logarithmic daily wages of the two groups (the higher wage group: $FC$; and the lower wage group: $NFC$)

$$W = E(Y_{FC}) - E(Y_{NFC}) \qquad 4$$

Here, $E = (Y)$ is the "expected value" of the log of daily wage of the workers in the $FC$ and $NFC$ group as indicated by the subscript.

Taking the logarithmic daily wage rate as a dependent variable, and demographically different characteristics of age, education, sector and region as predictors, the OLS wage equation is written as:

$$Y_k = X'_k \beta_k + \varepsilon_k \, ; \, E(\varepsilon_k) = 0, k \in \{NFC, FC\} \qquad 5$$

Where $X$ is a vector containing the predictors in subgroup $k$, $\beta$ is the slope parameter or the coefficient and $\varepsilon$ is the error term with zero mean and constant variance. The subscript $k$ denotes subgroups of the previously defined $FC$ and $NFC$ workers.

Since $E(\beta_k) = \beta_k$ and $E(\varepsilon_k) = 0$ as stated earlier, equation (5) gives us:

$$E(Y_k) = E(X_k)' \beta_k \qquad 6$$

Combining the equations (4) and (6) we get

$$W = E(X_{FC})'\beta_{FC} - E(X_{NFC})'\beta_{NFC} \qquad 7$$

To break down the overall difference into contributing components, equation (7) can be rewritten as:

$$W = [E(X_{FC}) - E(X_{NFC})]'\beta_{NFC} + E(X_{NFC})'(\beta_{FC} - \beta_{NFC}) + [E(X_{FC}) - E(X_{NFC})]'(\beta_{FC} - \beta_{NFC}) \qquad 8$$

Of which, the first term or the endowment component shows how much of the earnings differentials between two groups are caused by the differences in regressors. This term will measure the expected change in an individual $NFC$'s average earnings if he/she had $FC$ endowments (or observable characteristics of human capital). The endowment component is written as:

$$E = [E(X_{FC}) - E(X_{NFC})]'\beta_{NFC} \qquad 9$$

The second term or the coefficient component shows the contribution of differences in coefficients of the two groups. Simply put, a coefficient component will measure the "expected change" in $NFC$'s average earnings if he/she had an $FC$ coefficient. This is written as:

$$C = E(X_{NFC})'(\beta_{FC} - \beta_{NFC}) \qquad 10$$

And the third term or the interaction component considers the presence of endowments and coefficient differences at the same time or occurring simultaneously. Interaction is written as:

$$I = [E(X_{FC}) - E(X_{NFC})]'(\beta_{FC} - \beta_{NFC}) \qquad 11$$

Equation (8) is the Threefold decomposition and substituting (9), (10) and (11) we get

$$W = E + C + I \qquad 12$$

**6a. Empirical Results: Theil Index and Wage disparities 'between' and 'within' group**

To examine the overall inequality in wages we decompose the Theil index into between-group and within-group inequalities. The results are presented in Table 1. While employment shares of $NFC$ individuals (SC, ST and OBCs together) constitute about two-thirds of the total employment share, the mean wages are significantly lower than the other caste ($FC$) individuals. Male workers earn higher wages than female workers and this is observed across regions, sectors or even the nature of work. Not just the gender gaps are significant between the disadvantaged groups, we also note higher wage inequality among female workers, and this is true for both $FC$s and $NFC$s.

The $FC$ workers overall show a greater inequality which is 0.5515 and a lower inequality 0.3909 for $NFC$ workers. $FC$ females and $FC$ males both show higher inequalities as compared to $NFC$ females and males respectively. In the rural sector, the $FC$ workers show a higher inequality at 0.4359 and urban $FC$ workers show higher inequality at 0.4762.

Rural and $NFC$ workers contribute to more than 80 per cent employment share, yet they earn less than three fourth of what an average $FC$ worker earns. Within-group inequality is greater than between-group inequalities across socially disadvantaged individuals. The urban $NFC$ constitutes about 60 per cent as employment share, while the average wage for such individuals is significantly lower than the urban $FC$ individuals. Despite workers in the urban regions earning higher wages overall, there is a higher inequality prevalent in urban areas simultaneously and this is mainly attributable to "within-group inequality".

Among the $NFC$ individuals, the share of other backward caste individuals is higher and that corresponds with a higher wage as compared to the rest of the $NFC$ individuals. Inequalities observed for all social disadvantaged groups is much less than the forward group. And "within-group" proportions are greater than between groups in total wage across the social groups. For rural regions, we see that the between-group inequality contributes to less than 4 per cent of total inequality, this inequality share remains low for urban segregation of $NFC$ and $FC$ individuals at less than 8 per cent of total inequality.

**Table 1: Wage Gap Decomposition based on Caste (2011- 12)**

| Social Group | Employment Share (%) | Mean Wage | Gini index | Theil index | Within Group | | Between Group | |
|---|---|---|---|---|---|---|---|---|
| **Non-forward Caste** | 74.93 | 186.71 | 0.4487 | 0.3909 | | | | |
| **Forward Caste** | 25.07 | 368.27 | 0.5457 | 0.5515 | | | | |
| **Total Inequality** | | | 0.5059 | 0.5067 | 0.4547 | 89.74% | 0.0518 | 10.22% |
| | | | | | | | | |
| **Schedule Tribe** | 9.48 | 155.28 | 0.4717 | 0.4752 | | | | |
| **Schedule Caste** | 24.96 | 168.55 | 0.4192 | 0.3479 | | | | |
| **OBC** | 40.49 | 205.27 | 0.4529 | 0.3875 | | | | |
| **Others** | 25.07 | 368.27 | 0.5457 | 0.5515 | | | | |
| **Total Inequality** | | | 0.5059 | 0.5067 | 0.451 | 89.01% | 0.0557 | 10.99% |
| | | | | | | | | |
| **NFC Female** | 79.38 | 123.9 | 0.4546 | 0.4399 | | | | |
| **FC Female** | 20.62 | 299.81 | 0.605 | 0.6644 | | | | |
| **Total Inequality** | | | 0.5385 | 0.6108 | 0.5265 | 86.20% | 0.0842 | 13.79% |
| | | | | | | | | |
| **NFC Male** | 73.62 | 206.76 | 0.4295 | 0.358 | | | | |
| **FC Male** | 26.38 | 384.11 | 0.5302 | 0.5261 | | | | |
| **Total Inequality** | | | 0.4857 | 0.4687 | 0.4252 | 90.72% | 0.0435 | 9.28% |
| | | | | | | | | |
| **Rural NFC** | 81.89 | 144.54 | 0.3816 | 0.2907 | | | | |
| **Rural FC** | 18.11 | 211.25 | 0.4801 | 0.4359 | | | | |
| **Total Inequality** | | | 0.4103 | 0.338 | 0.3261 | 96.48% | 0.0124 | 3.67% |
| | | | | | | | | |
| **Urban NFC** | 61.47 | 295.42 | 0.4656 | 0.3814 | | | | |
| **Urban FC** | 38.54 | 511 | 0.515 | 0.4762 | | | | |
| **Total Inequality** | | | 0.5069 | 0.468 | 0.4307 | 92.03% | 0.0372 | 7.95% |

*Source: Authors own calculations based on NSS 2011-12 data.*

### 6b. Empirical Results: The Threefold Blinder Oaxaca Decomposition

From Table 2, education is significantly and positively related to the log of daily wages, however for the exception of below primary levels across castes, and a negative relationship (see *below primary* coefficient) for the forward castes (*others*) signifying perhaps both opportunities and motivations for forward castes to pursue further education. Higher coefficients for post-graduation across the castes indicate greater returns to higher education followed by graduation giving the highest returns for the forward castes indicating that these workers do not have to achieve the highest education to start getting better incomes.

In Table 2, other than education, we use personal characteristics such as age, gender, sector and region in our estimation. Comparing earnings, gender gaps for those belonging to *SC* and *Others* are higher, an SC or forward caste female worker earns 50 per cent less and the females in the rest of castes earn around 40 per cent less. Forward castes are observed to have an edge over the rest in urban areas as compared to rural (earning about 40 per cent higher), thus depicting a greater gap in the urban regions as compared to rural overall.

In Table 3a we show results for "original formulation of *E, C, U* and *D*" (Blinder-Oaxaca, 1973). Results indicate a high overall raw wage differential of 48.8 per cent. The raw wage differential is divided into three parts of which 43.1 per cent is attributable to endowment and a lower 6.1 per cent is attributable to discrimination (coefficient). The third unexplained "interaction term" is -0.4 per cent.

Results from Table 3b indicate a larger endowment component as compared to the discrimination component. The endowment component is 71.24 per cent as part of the total difference in the wage gaps. Nevertheless, discrimination explains 11.7 per cent lower wages and interaction explain about 17 per cent lower wages for *NFC* workers than that of the *FC* workers. Together the total attributable difference is close to 50 per cent (49.2) between the forward castes and non-forward castes workers and this is very large.

**Table 2: Earnings function OLS results in Regular salaried workers segregated by caste (2011- 12)**

| | Scheduled Castes | | | | Scheduled Tribes | | | | Other Backward Classes | | | | Others | | | |
|---|---|---|---|---|---|---|---|---|---|---|---|---|---|---|---|---|
| | coeff | std err | t-value | P>|t| | coeff | std err | t-value | P>|t| | coeff | std err | t-value | P>|t| | coeff | std err | t-value | P>|t| |
| age | 0.022625 | 0.002969 | 7.62 | 0.000 | 0.016663 | 0.003743 | 4.45 | 0.000 | 0.030794 | 0.002237 | 13.76 | 0.000 | 0.020773 | 0.002929 | 7.09 | 0.000 |
| agesq | -0.000202 | 0.000041 | -4.95 | 0.000 | -0.000129 | 0.000052 | -2.49 | 0.013 | -0.000321 | 0.000030 | -10.53 | 0.000 | -0.000117 | 0.000039 | -2.99 | 0.003 |
| below-primary | 0.023991 | 0.016200 | 1.48 | 0.139 | 0.012134 | 0.019158 | 0.63 | 0.527 | 0.027893 | 0.012372 | 2.25 | 0.024 | -0.030159 | 0.019481 | -1.55 | 0.122 |
| primary | 0.051652 | 0.014336 | 3.60 | 0.000 | -0.013699 | 0.018708 | -0.73 | 0.464 | 0.049378 | 0.011714 | 4.22 | 0.000 | 0.019409 | 0.016441 | 1.18 | 0.238 |
| secondary | 0.237148 | 0.018278 | 12.97 | 0.000 | 0.309578 | 0.026711 | 11.59 | 0.000 | 0.200062 | 0.012525 | 15.97 | 0.000 | 0.344197 | 0.015970 | 21.55 | 0.000 |
| highschool | 0.308350 | 0.024470 | 12.60 | 0.000 | 0.465028 | 0.032093 | 14.49 | 0.000 | 0.347976 | 0.016451 | 21.15 | 0.000 | 0.486816 | 0.018480 | 26.34 | 0.000 |
| grad | 0.730589 | 0.027823 | 26.26 | 0.000 | 0.780784 | 0.034458 | 22.66 | 0.000 | 0.844158 | 0.016221 | 52.04 | 0.000 | 1.065509 | 0.015611 | 68.25 | 0.000 |
| diploma | 0.680594 | 0.048218 | 14.12 | 0.000 | 1.030249 | 0.066691 | 15.45 | 0.000 | 0.737743 | 0.024044 | 30.68 | 0.000 | 0.924431 | 0.027240 | 33.94 | 0.000 |
| post-grad | 1.074412 | 0.043029 | 24.97 | 0.000 | 1.109383 | 0.059195 | 18.74 | 0.000 | 1.084990 | 0.022151 | 48.98 | 0.000 | 1.389023 | 0.019720 | 70.44 | 0.000 |
| Gender:male Base_Female | 0.522808 | 0.011777 | 44.39 | 0.000 | 0.396763 | 0.013809 | 28.73 | 0.000 | 0.516694 | 0.009005 | 57.38 | 0.000 | 0.529653 | 0.012224 | 43.33 | 0.000 |
| Sector:public Base_private | 0.591039 | 0.017786 | 33.23 | 0.000 | 0.645138 | 0.021927 | 29.42 | 0.000 | 0.513607 | 0.012850 | 39.97 | 0.000 | 0.508208 | 0.013457 | 37.77 | 0.000 |
| Region:urban Base_rural | 0.290183 | 0.012279 | 23.63 | 0.000 | 0.327365 | 0.018131 | 18.06 | 0.000 | 0.302432 | 0.008350 | 36.22 | 0.000 | 0.405578 | 0.010129 | 40.04 | 0.000 |
| _cons | 3.676148 | 0.052245 | 70.36 | 0.000 | 3.743405 | 0.065049 | 57.55 | 0.000 | 3.598352 | 0.039654 | 90.74 | 0.000 | 3.628384 | 0.053193 | 68.21 | 0.000 |
| R-squared | 0.3694 | | | | 0.4218 | | | | 0.4298 | | | | 0.5555 | | | |
| Adj- R2 | 0.3689 | | | | 0.4211 | | | | 0.4295 | | | | 0.5553 | | | |
| Observations | 14,144 | | | | 9,906 | | | | 25,892 | | | | 20,125 | | | |

*Source: Own calculations using NSSO microdata 68th round*
*Notes: p>0.10 = insignificant variable ; 0.01 < p < 0.05 = significant at 90 per cent level of confidence; 0.01 < p < 0.05 = significant at 95 per cent; p < 0.01 = significant at 99 per cent level of confidence.*

**Table 3a: Summary of The Threefold Blinder-Oaxaca Decomposition Results (as %)**

| Components of Decomposition | NFC vs FC |
|---|---|
| Total differential: | 49.2 |
|   - attributable to endowments (E): | 43.1 |
|   - attributable to coefficients (C): | 6.1 |
| Shift coefficient (U): | -0.4 |
| Raw differential (R) {E+C+U}: | 48.8 |
| Adjusted differential (D) {C+U}: | 5.7 |
| **Endowments as % total (E/R):** | **88.3** |
| **Discrimination as % total (D/R):** | **11.7** |

**Table 3b: The Threefold Blinder-Oaxaca Decomposition Results Components as a percentage of Total Difference**

| Components of Decomposition | NFC vs FC | % |
|---|---|---|
| Due to endowment (E) | 0.3475596 | 71.24 |
| Due to coefficients (C) | 0.0571873 | 11.72 |
| Due to interaction (I) | 0.0830914 | 17.03 |
| Gross Wage Differential (W = E+C+I) | 0.4878383 | 100.00 |

*Source: Own calculations based on NSS microdata 68th round*

Table 4 shows the relative contribution that each independent variable has on the wage gap. The results show that of the total difference in wages, how much is attributed to endowments and how much is attributed to differences in rewards[5]. Looking at levels of education, we see that except for below primary and primary level, all other (higher levels) favour forward caste workers.

Discrimination effect as part of the total difference in wages is stronger at below primary, primary and secondary levels as compared to the endowment effect. Moving up the educational levels reduces discrimination significantly. At the secondary level, we see that the total difference in wage between $FC$ and $NFC$ is 4.35 per cent, of which 2.34 per cent is due to discrimination and 1.37 per cent is attributable to the endowment. At the graduate level, the total difference is significantly higher at 28.10 per cent, and the discrimination component of the total is reduced. A similar pattern with effect to earnings differential and favourable treatment towards $FC$ is noted at the postgraduate level.

---

[5] 'Rewards' has been used in standard literature to show discrimination as a component of differential in Blinder Oaxaca decomposition.

**Table 4: Earnings Gap Three Fold decomposition (Blinder-Oaxaca) between NFC and FC using different variables**

|  | Endowments | % | Coefficients | % | Interaction | % | Total Difference |
|---|---|---|---|---|---|---|---|
| age | 0.022625 | 4.64 | -0.191533 | -39.26 | -0.004744 | -0.97 | -35.60 |
| agesq | -0.015925 | -3.26 | 0.181389 | 37.18 | 0.008575 | 1.76 | 35.68 |
| below-primary | -0.001053 | -0.22 | -0.006444 | -1.32 | 0.002327 | 0.48 | -1.06 |
| primary | -0.001220 | -0.25 | -0.003415 | -0.70 | 0.000670 | 0.14 | -0.81 |
| secondary | 0.006667 | 1.37 | 0.011426 | 2.34 | 0.003233 | 0.66 | 4.37 |
| highschool | 0.012214 | 2.50 | 0.007159 | 1.47 | 0.004437 | 0.91 | 4.88 |
| grad | 0.095075 | 19.49 | 0.013709 | 2.81 | 0.028306 | 5.80 | 28.10 |
| diploma | 0.011828 | 2.42 | 0.003173 | 0.65 | 0.002589 | 0.53 | 3.61 |
| post-grad | 0.065892 | 13.51 | 0.007206 | 1.48 | 0.017654 | 3.62 | 18.60 |
| male | 0.027385 | 5.61 | 0.017945 | 3.68 | 0.001281 | 0.26 | 9.55 |
| public | 0.047056 | 9.65 | -0.004398 | -0.90 | -0.003311 | -0.68 | 8.07 |
| urban | 0.077018 | 15.79 | 0.025254 | 5.18 | 0.022074 | 4.52 | 25.49 |
| constant | 0.000000 | 0.00 | -0.004283 | -0.88 | 0.000000 | 0.00 | -0.88 |
| **Subtotal** | 0.347560 | 71.24 | 0.057187 | 11.72 | 0.083091 | 17.03 | 100.00 |

*Source: Own calculations based on microdata from NSS 68th Round*

After education, the wage differentials are substantially greater for the urban area and favour the $FC$, showing a more pronounced wage gap in the urban area as compared to rural ones. By taking the wage structure of the $FC$, we see that 15.79 per cent of the total wage difference in urban regions is attributable to characteristics (or endowments) and 5.18 per cent is attributable to discrimination. An unexplained part of the wage differential is 4.52 per cent. Results in Table 4 also show a positive number for the public (8.07 per cent) indicating the presence of comparatively smaller discrimination against the $NFC$ in the public sector. The adjusted differential of 1.58 per cent shows a minuscule earning advantage favouring the disadvantaged workers in the public sector. The exceptions are the public sector where the discrimination component is negative and favours the $NFC$ and lower levels of education, where both endowments and discrimination components are negligible but favour the $NFC$.

A similar pattern is observed for the gender divide in wage differences, using the wage structure of the $FC$, a total differential in the wage gap between males and females is 9.55 per cent and favours the forward caste males. It is important to mention here that since our data does not fully account for differences in human capital, it will not be correct to assume that the full unexplained component is discriminatory even though in most variables, discrimination is less than endowment. Many women, for example, are excluded from the labour force due to caring and other "household obligations" (Kingdon 1998, Agarwal 2013).

**Discussion and Policy Implications**

Our research question approached wage disparities using two techniques. Through the first technique of Theil Index, we find that $NFC$ female workers face the greatest discrimination and this is evident from the higher Theil index found for this group (Deshpande, 2007 & Lama, 2018). However, comparing between the female $FC$ and $NFC$ workers, higher inequalities are found for the females belonging to the $FC$ groups. Interestingly, for all divisions, $FC$ shows greater within-group inequalities as compared to the $NFC$. Notwithstanding those standard deviations are seen high even in the $FC$ groups, which means that not everyone in this class is better off. The World Bank study reports that inequality within the sub-castes could be the main factor in economic inequality (Alvaredo et al., 2018). Despite the gap between the year of study in this paper and the report, findings have remained very similar showing the slow-moving rate of reduction in caste-related discrimination.

The Mincerian earnings function shows a rising premium to skill among the $NFC$ workers and this seems to be the trend post liberalisation of the Indian economy. However, the gaps between the castes are also attributed to increased wage inequalities more in urban areas as compared to rural areas. A substantial amount of labour market discrimination is found and it is also observed that to increase their standard of livings, the scheduled castes and scheduled tribes must move to a much higher level of education as compared to the forward caste workers and the other backward caste workers.

Comparing the levels of general education, higher education is shown to favour the forward caste while the lower levels pre-primary and primary shows favour towards the non-forward caste workers. This is an important observation, that strengthens our argument that the endogenous work division, which is traditionally fabricated into the societal and economic structure, is very much existent and even domination the market discrimination against the socially backward and disadvantaged groups. Women, at all levels and in all social and sectoral divisions are in a disadvantaged situation.

The decomposition results following the Oaxaca-Blinder approach allowed us to identify endowment, discrimination and interaction components of wage differentials. Even though the endowment component is larger than discrimination, the magnitude of discrimination cannot be ignored. A large endowment difference could imply pre-market discrimination with respect to human capital investments in education, health and nutrition and therefore becomes critical in explaining earning differentials than labour market discrimination. These pre-market discriminations have deprived those belonging to $NFC$ groups and policies to increase the endowment of not just physical capital (like providing of assets in the form of cattle, land, irrigation, wells and raw materials) but also human capital in forms of quality and affordable education and healthcare to begin with.

Our findings suggest that despite education continuing to be seen as a significant and positive investment for both disadvantaged and forward classes, the returns are higher at middle and graduate levels for the disadvantaged sections while they are the highest at the postgraduate level for the forward classes (Attewell & Madheswaran, 2007). To add to this, the returns are declining while going from graduate to a diploma level for all the three disadvantaged classes, showing a lack of participation of such sections in diploma courses. Recently, returns to education are seen the highest for postgraduate diploma courses and the lower levels of disadvantaged class participation suggests important policy implications. Policymakers should first invest in basic quality education and simultaneously expand post-graduate diploma opportunities, subsequently increasing the participation in the labour force for the traditionally disadvantaged sections in disciplines and occupations where the forward castes have long dominated (Awasthi et al., 2015; Borooah et al., 2014).

Providing educational empowerment in forms of pre-matric, post-matric scholarships, fellowships, free coaching services, to support the children of marginalised groups, will ensure quality education and lower the incidence of dropouts are strong policy tools. Closing the income inequality gaps will also call for entrepreneurship programs, skill training, refinancing loans, credit facilities with the aim to encourage entrepreneurship will result in not just the creation of jobs but also help bring such sections into the mainstream of growth and development discourse. Several educational and economic initiatives are taken by the Ministry of Social Justice and Empowerment, such as the Post Matric Scholarship for Scheduled Caste Students (PMS-SC) that aims to assist students belonging to scheduled castes by providing scholarships, fellowships and free coaching services. Policies such as the Credit Enhancement Guarantee Scheme for Scheduled Castes, SCDCs, NSFDC, SCSP are also initiated.

A newly developing trend where returns to education are seen higher at the tertiary level does have important policy implications for the deprived sections as the demand for tertiary education rises, so does the requirements of higher education where the enrollment of such sections have remained very low. The gross enrollment ratios (GER) for SCs at primary and upper primary levels are over 100, however, they start dipping while moving up the education ladder, falling to 82.7 at the secondary level, and a dismal 19.1 for higher education (Social Welfare Statistics, 2018). Policymakers should however continue to improve access to quality primary and secondary education that is an important prerequisite for entering higher education.

Narrowing the endowment component is crucial to the "functioning of any democratic government" (Chakraborty, 2016), this is an important issue because the pre-market disparities in human capital including quality education, skill development and training manifest as employer's bias against the disadvantaged groups leading to discrimination based on other observable characteristics such as age, gender, disability and at times region (Charles & Guryan, 2007). Perceptions can be very challenging to change, nevertheless, continuous investments in high-quality primary education initially and skill development and training, later on, can gradually improve commonly held perceptions against SCs, STs and OBCs.

Arguments against reservations that are based on the so-called ineffectiveness and inefficiency of such policies are neither empirically documented nor supported. Such policies do have incentives for the non-forward caste individuals to access better and higher education, given that in absence of which most such individuals would have not pursued. The reservation policies can be seen as a system that allocates resources such as seats in colleges and government jobs (Munshi, 2019). Despite reservations policies seen as in fact redistributing opportunities (Bertrand et al., 2010, Munshi, 2019) and a range of reservations for SC, STs and OBCs in civil posts and services, the proportion covered by reservations remains minuscule. This could be due to the higher employment of such workers in informal, non-reserved non-government jobs.

In rural areas where inequalities are observed lower as compared to the urban areas, the opportunities for growth and increasing incomes are restrained and limited. The rural labour market has traditionally been caste-based (Saha & Verick 2016; Carswell, 2013). Wage labourers and farmers belonging to the disadvantaged castes face discrimination in the sense that their goods and services are less demanded by the traditionally privileged castes, or discrimination is found while buying raw materials and inputs, obviously affecting the wage of this group. Special emphasis is already given for the inclusion of scheduled castes and tribes in providing awareness on MGNREGA and additional provisions must be made to NFC individuals for land development, provision of irrigation facility, plantation and horticulture, among others.

UNDP's Global Multidimensional Poverty Index 2021 shows prevalent widespread poverty predominantly among the scheduled castes (SCs) and tribes (STs); with the poverty ratio highest among the SCs at 50.6 per cent, subsequently for SCs and OBCs are high at 33.3 per cent and 27.2 per cent respectively. This is in contrast to 15.6 per cent for the forward caste. Despite the results from this paper highlighting a wide gap between the castes based on data from 2011-12, the present situation with widely prevalent poverty and discrimination among the disadvantaged social groups is not very different. As much as reducing visible disparities in education and income, bringing down discrimination remains a challenge. Caste-based wage discrimination can counteract the development process. In the past, caste-related violence has reinstated the traditional differences between the so-called higher caste individuals and the socially disadvantaged individuals, acquiring new vigour and turning into a violent and fierce struggle for power in our incessant hierarchical society. Marginalised and backward castes need to be brought into the mainstream of the ongoing development process to achieve holistic growth. An important policy implication is that there is a need to expand economic initiatives based on an area-specific approach rather than a caste-based approach, therefore targeting vulnerable populations in deprived areas regardless of their castes can narrow large gaps in inequalities and lessen communal tensions or caste-wars as well.

# Appendix 1. Descriptive Statistics of selected variables

| Variables | Description of the Variables | Scheduled Caste | | Scheduled Tribe | | Other Backward Caste | | Others | |
|---|---|---|---|---|---|---|---|---|---|
| | | Mean | Std. Dev | Mean | Std. Dev | Mean | Std. Dev | Mean | Std. Dev |
| lwage | Logarithm of daily wage (in rupees) | 4.820057 | 0.7475881 | 4.659777 | 0.8025349 | 4.974421 | 0.7933116 | 5.371033 | 1.006541 |
| Age | Age in years | 34.67003 | 10.82573 | 34.2806 | 10.6227 | 34.91094 | 10.79073 | 35.61171 | 10.87655 |
| agesq | Age squared | 1319.199 | 787.6599 | 1287.99 | 768.6686 | 1335.209 | 792.6708 | 1386.487 | 809.9557 |
| <Primary | if completed below primary education=1; 0 otherwise | 0.1173098 | 0.3218004 | 0.1290705 | 0.3352949 | 0.113963 | 0.3177724 | 0.0747498 | 0.2629937 |
| Primary | if completed primary education=1; 0 otherwise | 0.1632851 | 0.3696387 | 0.1419318 | 0.3489978 | 0.1330588 | 0.3396448 | 0.1159301 | 0.320149 |
| Secondary | if completed secondary education=1; 0 otherwise | 0.092595 | 0.2898743 | 0.0631702 | 0.2432812 | 0.1162562 | 0.320538 | 0.1304215 | 0.3367749 |
| HSC | if completed higher secondary=1; 0 otherwise | 0.0479179 | 0.2136001 | 0.0452674 | 0.2079004 | 0.0619876 | 0.2411376 | 0.08939 | 0.285313 |
| Grad | if completed graduation=1; 0 otherwise | 0.0380705 | 0.1913732 | 0.0435283 | 0.2040534 | 0.0701238 | 0.2553605 | 0.1718776 | 0.3772834 |
| Diploma | if completed diploma/ certificate course=1; 0 otherwise | 0.0111947 | 0.1052149 | 0.008887 | 0.0938556 | 0.0263876 | 0.1602881 | 0.0846967 | 0.1830432 |
| Postgrad | if completed post-graduation=1; 0 otherwise | 0.0145648 | 0.1198068 | 0.0121057 | 0.1093635 | 0.0336168 | 0.1802443 | 0.0347078 | 0.2784368 |
| Gender-Male | If male =1; 0 otherwise | 0.7544501 | 0.430428 | 0.6979318 | 0.4591779 | 0.7743006 | 0.4180501 | 0.8121484 | 0.3906033 |
| Sector-Public | If private sector =1; 0 otherwise | 0.1052055 | 0.3068288 | 0.1314364 | 0.3378941 | 0.1159791 | 0.320206 | 0.2004234 | 0.4003271 |
| Region-Urban | If rural region =1; 0 otherwise | 0.2425139 | 0.4286186 | 0.1601833 | 0.3667945 | 0.3302627 | 0.4703167 | 0.5238378 | 0.4994438 |

| Occupation | | | | | | | | | |
|---|---|---|---|---|---|---|---|---|---|
| NCO_1 | Base 1; 0 otherwise | 0.0160774 | 0.1257777 | 0.0154529 | 0.1233519 | 0.029184 | 0.1683252 | 0.0733345 | 0.2606912 |
| NCO_2 | Base 1; 0 otherwise | 0.0192926 | 0.1375561 | 0.0172562 | 0.1302312 | 0.0394305 | 0.1946206 | 0.0959298 | 0.2945023 |
| NCO_3 | Base 1; 0 otherwise | 0.0271227 | 0.1624467 | 0.0366705 | 0.187961 | 0.0388395 | 0.1932161 | 0.0798028 | 0.2709943 |
| NCO_4 | Base 1; 0 otherwise | 0.0241085 | 0.1533916 | 0.0227741 | 0.14919 | 0.0310898 | 0.1735639 | 0.0694625 | 0.2542454 |
| NCO_5 | Base 1; 0 otherwise | 0.0411633 | 0.1986748 | 0.0411049 | 0.1985428 | 0.0730739 | 0.2602627 | 0.0952546 | 0.2935736 |
| NCO_6 | Base 1; 0 otherwise | 0.0593044 | 0.236202 | 0.1243092 | 0.3299506 | 0.0751737 | 0.2636765 | 0.0587028 | 0.2350735 |
| NCO_7 | Base 1; 0 otherwise | 0.1552232 | 0.3621301 | 0.0920895 | 0.2891668 | 0.1781299 | 0.3826295 | 0.146478 | 0.3535936 |
| NCO_8 | Base 1; 0 otherwise | 0.058884 | 0.2354159 | 0.0347777 | 0.1832257 | 0.0849231 | 0.2787726 | 0.1073212 | 0.3095289 |
| NCO_9 | Base 1; 0 otherwise | 0.5959554 | 0.4907235 | 0.6129229 | 0.4871061 | 0.4485694 | 0.4973575 | 0.2674765 | 0.442654 |

Source: Own calculations based on NSS microdata 68th round

Notes: Standard deviations are not reported for dummy variables.

1. NCO Classifications (2004) are as follows: NCO1: Legislators, Senior Officials and Managers; NCO2: Professionals; NCO3: Technicians & Associates Professionals; NCO4: Clerks; NCO5: Service Workers & Shop & Market Sales Workers; NCO6: Skilled Agricultural and Fishery Workers; NCO7: Craft and Related Trades Worker; NCO8: Plant and Machinery Operators and Assemblers; NCO9: Elementary Occupations.

# References

Agrawal, T. (2013). Gender and caste-based wage discrimination in India: some recent evidence. *Journal for Labour Market Research*, *47*(4), 329–340. https://doi.org/10.1007/s12651-013-0152-z

Akita, Takahiro . (2003). Decomposing Regional Income Inequality in China and Indonesia Using Two-stage Nested Theil Decomposition Method. *Annals of Regional Science* 37(1), 55–77.

Allison, Paul D. (1978). Measures of Inequality. *American Sociological Review,* 43(6), 865–80

Alvaredo, F., Chance, L., Piketty, T., Saez, E., & Zucman, G. (2018). World Inequality Report 2018. In *Wid.world*. https://wir2018.wid.world/

Awasthi, I., Shrivastav, P., & Kumar. (2015). *Munich Personal RePEc Archive Inequalities in Economic and Educational Status in Social Groups in India: Evidences from Village Study in Uttar Pradesh*. https://mpra.ub.uni-muenchen.de/66441/1/MPRA_paper_66441.pdf

Ben Jann, (2008). A Stata implementation of the Blinder-Oaxaca decomposition. No 5. *ETH Zurich Sociology Working Papers*. ETH Zurich, Chair of Sociology.

Ben Jann, 2008. The Blinder–Oaxaca decomposition for linear regression models. *Stata Journal,* 8(4), 453-479.

Bertrand, M., Hanna, R., & Mullainathan, S. (2010). Affirmative action in education: Evidence from engineering college admissions in India. Journal of Public Economics, 94(1-2), 16–29. https://doi.org/10.1016/j.jpubeco.2009.11.003

Bharti, N. (2019). *Local Level Land Inequality in India*. https://www.isid.ac.in/~epu/acegd2019/papers/NitinBharti.pdf

Bhuyan, B., Sahoo, B. K., & Suar, D. (2018, May 23). *A Quantile Decomposition of Household's Food Security in India by Caste*. Papers.ssrn.com. https://ssrn.com/abstract=3183902

Banerjee, B., & Knight, J. (1985). Caste discrimination in the Indian urban labour market. *Journal of Development Economics*, 17(3), 277-307.

Borooah, V. K., & Iyer, S. (2005). Vidya, Veda, and Varna: The influence of religion and caste on education in rural India. *The Journal of Development Studies*, *41*(8), 1369–1404. https://doi.org/https://doi.org/10.1080/00220380500186960

Borooah, V. K., Diwakar, D., Mishra, V. K., Naik, A. K., & Sabharwal, N. S. (2014). Caste, inequality, and poverty in India: a re-assessment. *Development Studies Research*, *1*(1), 279–294. https://doi.org/10.1080/21665095.2014.967877

Bourguignon, F., H.G. Ferreira Francisco, and M. Menéndez, "Inequality of Opportunity in Brazil," Review of Income and Wealth, 53(4), 585-618, 2007

Carswell, G. (2013). Dalits and local labour markets in rural India: experiences from the Tiruppur textile region in Tamil Nadu. *Transactions of the Institute of British Geographers*, *38*(2), 325–338. http://www.jstor.org/stable/24582477

Chakraborty, N. & Majumder, A. (2016). Occupational Segregation and wage differential between males and females in India. *Sarvekshna, 100th issue*.

Charles-Coll, Jorge A. (2011). Understanding Income Inequality: Concept, Causes and Measurement. *International Journal of Economics and Management Sciences.* 1(1), 17–28.

Charles, K. K., & Guryan, J. (2007). *Prejudice and The Economics of Discrimination*. National Bureau of Economic Research Working Paper Series. http://www.nber.org/papers/w13661

Cotton, J. (1988). On decomposition of wage differentials. *Review of Economics and Statistics,* 70(2), 236 - 243.

Das, M. B., & Dutta, P. V. (2007). Does Caste Matter for Wages in Indian Labour Market? https://conference.iza.org/conference_files/worldb2008/dutta_p4261.pdf

Das, P. (2018). Wage Gap and Employment Status in Indian Labour Market. *World Journal of Applied Economics*, *4*(2), 117-134. https://doi.org/10.22440/wjae.4.2.4

Deshpande, A. (2007). Overlapping Identities under Liberalization: Gender and Caste in India. *Economic Development and Cultural Change*, *55*(4), 735–760. https://doi.org/10.1086/516763

Deshpande, A., & Sharma, S. (2016). Disadvantage and discrimination in self-employment: caste gaps in earnings in Indian small businesses. *Small Business Economics*, *46*(2), 325–346.

Faruk, A. (2019). Analysing the glass ceiling and sticky floor effects in Bangladesh: evidence, extent and elements. *SN Business & Economics*, *1*(9). https://doi.org/10.1007/s43546-021-00123-z

Handbook on Social Welfare Statistics. (2018). *Government of India. Ministry of Social Justice & Empowerment*. Plan Division, New Delhi. Available at https://socialjustice.nic.in/

Jann, B., & Zurich, E. (2008). *A Stata implementation of the Blinder-Oaxaca decomposition A Stata implementation of the Blinder-Oaxaca decomposition*. https://core.ac.uk/download/pdf/6442665.pdf

Jann, B. (2008). A Stata implementation of the Blinder-Oaxaca decomposition. No 5. *ETH Zurich Sociology Working Papers*. ETH Zurich, Chair of Sociology.

Jann, B. (2008). The Blinder–Oaxaca decomposition for linear regression models. *Stata Journal,* 8(4), 453-479

Jann, B., & Zurich, E. (2008). A Stata implementation of the Blinder-Oaxaca decomposition. https://core.ac.uk/download/pdf/6442665.pdf

Jenkins, Stephen P. (1995). Accounting for Inequality Trends: Decomposition Analysis for the UK. *Economica* 62(1), 29–64.

Jenkins, Stephen P. (1999). INEQDECO: Stata Module to Calculate Inequality Indices with Decomposition by Subgroup. Revised May 26, 2008. *Statistical Software Components S366002.* Boston: Boston College Department of Economics.

Jayadev, Arjun and Sanjay G. Reddy (2011). "Inequalities and Identities". SSRN Electronic Journal

Karki, M., & Bohara, A. K. (2014). Evidence of Earnings Inequality Based on Caste in Nepal. *The Developing Economies*, *52*(3), 262–286. https://doi.org/10.1111/deve.12049

Kijima, Y. (2006). Caste and Tribe Inequality: Evidence from India, 1983-1999. *Economic Development and Cultural Change*, *54*(2), 369–404.
25

Kingdon, G. G. (1998). Does the labour market explain lower female schooling in India? *Journal of Development Studies*, *35*(1), 39–65. https://doi.org/10.1080/00220389808422554

Klasen, S. (2018). The impact of gender inequality on economic performance in developing countries. Discussion Papers, No. 244, Georg-August-Universität Göttingen. *Courant Research Centre - Poverty, Equity and Growth (CRC-PEG)*, Göttingen

Khanna, S. (2012). Gender Wage Discrimination in India: Glass Ceiling or Sticky Floor? *SSRN Electronic Journal, Working Paper No. 214*. https://doi.org/10.2139/ssrn.2115074

Kuriakose, F., & Iyer, D. K. (2020). Job Polarisation in India: Structural Causes and Policy Implications. *The Indian Journal of Labour Economics*, *63*(2), 247–266. https://doi.org/10.1007/s41027-020-00216-7

Kumar, M., & Pandey, S. (2021). Wage Gap Between Formal and Informal Regular Workers in India: Evidence from the National Sample Survey. *Global Journal of Emerging Market Economies*, *13*(1), 104–121. https://doi.org/10.1177/0974910121989458

Lama, S. & Majumder, R. (2018). Gender inequality in wage and employment in Indian labour market. *Journal of Academic Research in Economics*, 10(3), 482-500.

Liao, Tim. (2016). THEILDECO: Stata module to produce refined Theil index decomposition by group and quantile. Statistical Software Components S458187, Boston College Department of Economics.

Mainali, R., Jafarey, S., & Montes-Rojas, G. (2016). Earnings and Caste: An Evaluation of Caste Wage Differentials in the Nepalese Labour Market. *The Journal of Development Studies*, *53*(3), 396–421. https://doi.org/10.1080/00220388.2016.1189535

Madheswaran, S., & Singhari, S. (2016). Social exclusion and caste discrimination in public and private sectors in India: A decomposition analysis. *The Indian Journal of Labour Economics*, *59*(2), 175–201. https://doi.org/10.1007/s41027-017-0053-8

Madheswaran, S. & Shroff, S. (2000). Education, employment and earnings for scientific and technical workforce in India: Gender issues. *Indian Journal of Labour Economics*, 43(1), 121-37.

Madheswaran, S., Attewell, P. (2007). Caste discrimination in the Indian urban labour market: Evidence from the National Sample Survey. *Economic and Political Weekly*, 42(41), 4146–4153.

Montes-Rojas, G., Siga, L., & Mainali, R. (2017). Mean and quantile regression Oaxaca-Blinder decompositions with an application to caste discrimination. *The Journal of Economic Inequality*, 15(3), 245-255.

Mukherjee, D., & Majumder, R. (2011). Occupational pattern, wage rates and earning disparities in India: A decomposition analysis. *Indian Economic Review*, 131-152.

Munshi, K. (2019). Caste and the Indian Economy. *Journal of Economic Literature*, *57*(4), 781–834. https://doi.org/10.1257/jel.20171307

Newmark, D. (1988): Employer Discriminatory Behaviour and the Estimation of Wage Behaviour, *Journal of Human Resources*, 23, 279-95

Oaxaca, R. (1973). Male female differentials in urban labour markets. *International Economic Review,* 14, 693-709.

Oaxaca, R. L., & Ransom, M. R. (1994). On discrimination and the decomposition of wage differentials. *Journal of Econometrics*, *61*(1), 5–21. https://doi.org/10.1016/0304-4076(94)90074-4

Reza, A. G., Gupta, P., Mishra, T., & Parhi, M. (2018). Wage differential between caste groups: are younger and older cohorts different? - *LSE Research Online. ISSN 0264-9993*, *74*, 10–23. https://doi.org/http://eprints.lse.ac.uk/90510/1/Arabsheibani__wage-differential.pdf

Saha, P., & Verick, S. (2016). *ILO Asia - Pacific Working Paper Series: State of rural labour markets in India*. https://www.ilo.org/wcmsp5/groups/public/---asia/---ro-bangkok/---sro-new_delhi/documents/publication/wcms_501310.pdf

Sangwan, N. (2020). *Greenwich Papers in Political Economy 3000 Years of Discrimination and Counting: How Caste Still Matters in the Indian Credit Sector*. https://docs.gre.ac.uk/__data/assets/pdf_file/0021/198300/Sangwan-2020-3000-Years-of-Discrimination-and-Counting.pdf

Sharma, S. (2017). Effect of Endowments on Gender Wage Differentials: A Decomposition Analysis for Indian Labour Market. *Economic Affairs*, *62*(4), 609. https://doi.org/10.5958/0976-4666.2017.00074.2

Sikdar, S. (2019). Rate of Return to Education in India: Some Insights. In *Working Paper No. 270*. https://www.nipfp.org.in/media/medialibrary/2019/06/WP_270_2019.pdf

Somasree Poddar & Ishita Mukhopadhyay, (2019). "Gender Wage Gap: Some Recent Evidences from India," *Journal of Quantitative Economics, Springer; The Indian Econometric Society (TIES), vol. 17(1)*, pages 121-151, March.

Tagade, N., Naik, A. K., & Thorat, S. (2018). Wealth Ownership and Inequality in India: A Socio-religious Analysis. *Journal of Social Inclusion Studies, 4(2)*, 196-213.

Theil, H. (1967). Economics and information theory. *Studies in mathematical and managerial economics*, v.7, No. 04; HB74. M3, T4.

Thorat, S., & Madheswaran, S. (2018). Graded caste inequality and poverty: evidence on role of economic discrimination. *Journal of Social Inclusion Studies,* 4(1), 3-29.

Thorat S. (2018, September 7). Scheduled Castes among worst sufferers of India's job problem. Hindustan Times; Hindustan Times. https://www.hindustantimes.com/india-news/scheduled-castes-among-worst-sufferers-of-india-s-job-problem/story-Qh0hyHy9UUTg1cIOpi5l2K.html

UNDP (2021). *Global Multidimensional Poverty Index 2021*, *Unmasking disparities by ethnicity, caste and gender,* UNDP

United Nations. Department of Economic and Social Affairs. (2020). *World Social Report 2020: Inequality in a Rapidly Changing World*. UN.